\documentclass{svproc}
\usepackage{url}
\usepackage[numbers,sort&compress,sectionbib]{natbib}% Citation support using natbib.sty
\usepackage{soul}
\usepackage{color,soul}
\usepackage{graphicx}
\usepackage{amsmath}
\usepackage{caption}
\usepackage{subfig}
\usepackage{xcolor,colortbl}
\usepackage{amsmath}
\usepackage{amsfonts}
\usepackage{steinmetz}
\usepackage{booktabs}

\begin{document}
\mainmatter              % start of a contribution

\title{Modeling Head-Neck Dynamics under Lateral Perturbations Using MPC to Mimic CNS postural stabilization strategy}
% %
\titlerunning{MPC-Based Modeling of Head-Neck Dynamics under Lateral Perturbations}  % abbreviated title (for running head)
\author{Chrysovalanto Messiou \and  Riender Happee \and  Georgios Papaioannou}
% %
\authorrunning{Messiou et al.} % abbreviated author list (for running head)
% %
% %%%% list of authors for the TOC (use if author list has to be modified)
\tocauthor{Chrysovalanto Messiou, Riender Happee and Georgios Papaioannou}
% %
\institute{Delft University of Technology, The Netherlands\\
 \email{c.messiou@tudelft.nl}}

\maketitle              % typeset the title of the contribution

\begin{abstract}
Automated vehicles will allow occupants to engage in non-driving tasks, but limited visual cues will make them vulnerable to unexpected movements. 
These unpredictable perturbations create a “surprise factor,” forcing the central nervous system to rely on compensatory postural adjustments, which are less effective, and are more likely to trigger sensory conflicts. 
Since the head is a key reference for sensory input (vestibular and vision), models accurately capturing head-neck postural stabilization are essential for assessing AV comfort. 
This study extends an existing model predictive control-based framework to simulate head-neck postural control under lateral perturbations. 
Experimental validation against human data demonstrates that the model can accurately reproduce dynamic responses during lateral trunk perturbations. 
The results show that muscle effort combined with partial somatosensory feedback provides the best overall dynamic fit without requiring corrective relative and global head orientation integrators for posture. 


\end{abstract}

% \textcolor{red}{
% Comments:
% The authors should more clearly define position T1 referred to in the current version of the paper
% For the work to be relevant to the conference, however, the relation between vehicle motion and driver motion needs to be clearer. In the proposed work, the vehicle appears entirely absent?
% }

\section{Introduction}

In automated vehicles (AVs), occupants are expected to take their eyes off the road and make use of their commute time \cite{yunus2024review}. 
The limited visual cues about the vehicle's upcoming motion hinder the occupants' anticipatory postural control in the presence of unexpected disturbances, leaving the central nervous system's (CNS) reliant on compensatory postural adjustments (CPA) \cite{papaioannou2025occupants}.
CPAs initiated by sensory feedback post-perturbation, involve reflexive and voluntary phases to restore balance.
When CPA becomes the sole mechanism due to the absence of accurate sensory predictions, a ``surprise factor" arises. 
This factor makes adjustments less efficient and more energy-intensive \cite{Santos1}, increases abrupt head motion, and triggers sensory conflict \cite{Kuiper2020}, assumed as the main cause of motion sickness in AVs. 

The postural stabilization of the head-neck system depends on integrating sensory information (visual, vestibular, somatosensory, and auditory inputs) with the CNS's ``memory" of prior movements, enabling sensory prediction via the neural store \cite{oman1991}.
Accurate sensory information and reliable CNS predictions regarding relative and global orientation allow the CNS to issue precise control actions. 
In fact, minimizing sensory conflict between predicted and sensed orientation signals during destabilizing perturbations is hypothesized as a core CNS objective \cite{ActiveInferenceFriston}.
However, few models attempt to capture sensory integration in relation to postural stabilization   \cite{Happee2023}. 
Meanwhile, existing neck models are often slow due to their biomechanical complexity \cite{Meyer2013,Correia2020,Correia2021} and rely on overly simplified control frameworks that do not account for the plausible CNS’s inference and belief processes. 
As a result, their ability to assess head-neck dynamics in dynamic driving scenarios is limited.

Messiou et al. \citep{Messiou2025} introduced the first model predictive control (MPC)-based postural control framework for the head-neck system. 
The model, developed in Simscape (MATLAB), represents head-neck dynamics as a simplified biomechanical multi-body system.
The MPC reflects CNS functionality by predicting future behavior through internal models, optimizing control inputs to minimize sensory conflict within biomechanical constraints. 
By incorporating sensory errors from human motion perception systems into the MPC cost function \cite{Messiou2025}, the authors model a plausible CNS inference strategy \cite{ActiveInferenceFriston}, assuming that the primary decision-making objective is to minimize the ``surprise factor"—the difference between predicted and actual sensory feedback.
The new model was validated for translational and rotational perturbations in the sagittal plane, achieving accurate predictions with a single parameter set.
This paper extends validation to lateral perturbations with eyes closed, covering the coronal plane.

\section{MPC Based Postural Control}

This section presents an overview of the framework as developed by Messiou et al. \cite{Messiou2025}, describes its improvements compared to the previous version, and demonstrates the high-level optimization function used to tune the MPC weights. 
In addition, we elaborate on the model posture and its effect on the dynamic response of the model.

\subsection{MPC based postural control framework}

\begin{figure}[h!]
  \centering \includegraphics[width=1\linewidth] {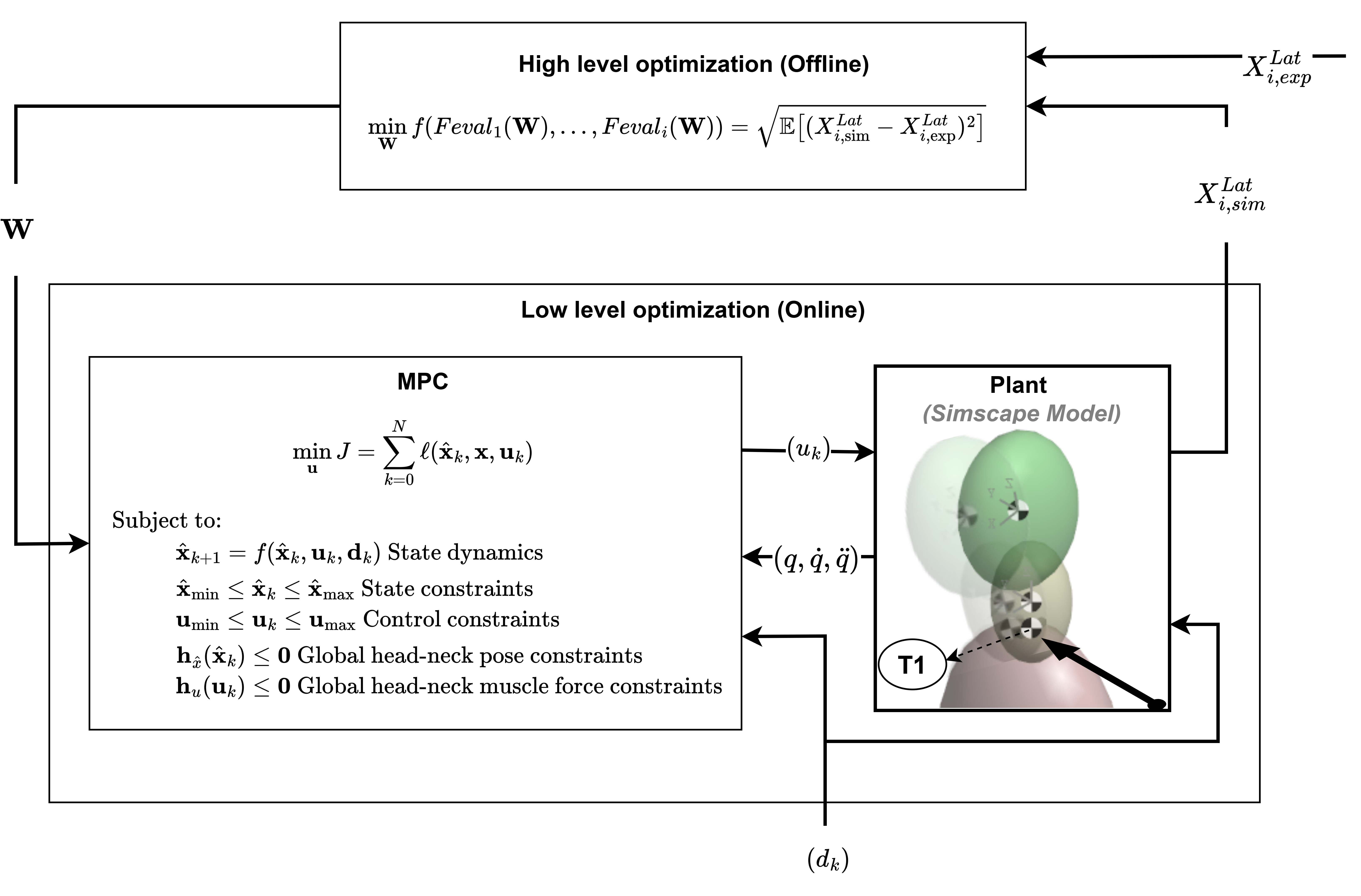}
  \caption{Block diagram of the postural stabilization framework. High-level optimization computes the weight vector ($W$) for the MPC cost function ($J$), by minimizing the RMSE between simulated ($X_{i,sim}^{Lat}$) and experimental ($X_{i,exp}^{Lat}$) human responses. Control inputs ($u_k$)  are generated in response to lateral disturbances ($d_k$) by minimizing the ``surprise factor" between actual sensory feedback (${q}, \dot{q}, \ddot{q}$) and predictions from the neural store (MPC prediction model). }
  \label{fig:MPC_Concept}
\end{figure}

The block diagram in Fig. \ref{fig:MPC_Concept}, illustrates briefly the postural stabilization framework using MPC \cite{Messiou2025}. 
Upon disturbance ($d_k$), the CNS receives sensory feedback via proprioceptive, vestibular, and visual pathways, derived here from the Simscape model's states ($q, \dot{q}, \ddot{q}$). 
The MPC predicts sensory feedback ($\hat{q}, \hat{\dot{q}}, \hat{\ddot{q}}$) using ODEs \cite{Messiou2025}, and minimizes sensory conflict, muscle effort, and adheres to biomechanical constraints, applying optimized motor commands to the plant ($u_k$).
The optimization process \cite{Messiou2025} tunes MPC weights using average head-neck responses from lateral perturbation datasets via MATLAB's \textit{multiga()} function, minimizing RMSE between experimental ($X_{i,exp}^{Lat}$) and simulated ($X_{i,sim}^{Lat}$) responses.
The tuning of the cost function weights through a high level optimization is illustrated in Fig. \ref{fig:MPC_Concept}.

\subsection{MPC configuration}

To improve computational performance and robustness in the coronal plane, several modifications were made to the MPC-based control framework \cite{Messiou2025}. 
Firstly, the yaw degree of freedom (DoF) of the lower neck joint was constrained (locked) in both the Simscape model and the ODE-based prediction model. 
This simplification was motivated by the observation that lower-neck yaw had minimal impact on the realism of head-neck dynamics compared to upper-neck yaw, while contributing significantly to model complexity. 
Moreover, the axial mobility of the human neck joints is concentrated in the upper neck \cite{panjabi2001}.

Secondly, the prediction horizon configuration was revised. 
The forecast horizon \(T_{H}\) was re-defined by a multi-collocation time interval approach, where \(T_{H}\) includes intermediate evaluation nodes (collocation points) within each interval (Eq. \ref{eq:forecast}).
This change allows longer prediction horizons without increasing numerical integration time steps.

\begin{equation}
T_{H} = \{ t_0 \} \cup \bigcup_{k=0}^{N-1} \left\{ t_k + T_{sp} \cdot \tau_j \;\middle|\; j = 1, \dots, d \right\} \cup \{ t_N \}
\label{eq:forecast}
\end{equation}

\noindent where \( {T}_{H} \) is the set of all forecast time points over the prediction horizon; \( t_0 \) is the initial time (current time); \( t_k = t_0 + k h \) is the beginning of the \(k\)-th interval; \( t_N = t_0 + T \) is the final prediction time; \( T_{sp} = \frac{T_{H}}{N}=40 \,ms \) is the length of each time interval; \( \tau_j \in (0, 1] \) is the normalized location of the \(j\)-th intermediate node (collocation point) withinl interval \(k\); \( N = \, 10 \) is the total number of time intervals in the prediction horizon and \( d = \, 4\) is the number of intermediate nodes (collocation points) per interval.
This enhanced both accuracy and real-time performance.
System dynamics were integrated at \(10 \, ms\).

Lastly, additional terms were incorporated into the cost function, including weights on roll and yaw angular velocities and muscle effort. 
Based on previous findings \cite{Messiou2025}, the configuration of partial somatosensory feedback and muscle effort was identified as the best overall MPC configuration for dynamic responses under anterior-posterior (AP) perturbations and was retained in the present work.

\subsection{High – Level optimization}

The high-level optimization procedure aims to tune the cost function weights of the MPC framework offline, ensuring that the model replicates experimentally observed head-neck responses during lateral perturbations. 
This process follows previous work \cite{Messiou2025}, where multi-objective genetic algorithms via MATLAB's \textit{multiga()} were applied to minimize the root mean square error (RMSE) between simulated and experimental signals. 
The optimization problem is defined as:

\begin{equation}
\min_{\mathbf{W}} f(\text{Feval}1(\mathbf{W}), ..., \text{Feval}i(\mathbf{W})) = \ \sqrt{\mathbb{E} \left[(X_{i,\text{sim}}^{S} - X_{i,\text{exp}}^{S})^2 \right]}
\end{equation}

\noindent where \( \mathbf{W} \) represents the vector of cost function weights, and each \( \text{Feval}_i \) corresponds to an objective function quantifying the error between simulated (\( X_{i,\text{sim}}^{S} \)) and experimental signals (\( X_{i,\text{exp}}^{S} \)) for the lateral perturbation scenario (\( S = Lat \)).  
The objective functions considered include time-domain metrics— global angular head position about x- and z-axis (`roll', `yaw'); y-axis global translation head position (`y'); global angular head velocity about x- and z-axis (`wroll', `wyaw'); y-axis global translational head velocity (`vy'). 
Additionally frequency response functions from T1 (first thoracic vertebra) to the head were used— y-axis velocity gain (\( \|f_{vy}\| \)) and phase (\( \phase f_{vy} \)); roll rate of orientation gain (\( \|f_{wroll}\| \)) and phase (\( \phase f_{wroll} \)); yaw rate of orientation gain (\( \|f_{wyaw}\| \)) and phase (\( \phase f_{wyaw} \)).

\section{Experimental Dataset} \label{Vehicle Model}

The experimental dataset was obtained from a study by Forbes et al. \cite{forbes2014}, where participants seated on a motion platform without a headrest experienced pseudorandom multisine perturbations applied to the seat. The torso was fixed to the seat with a harness, approximating the primary perturbation point at T1 (Fig. \ref{fig:HeadNeckModel}).
Two conditions were tested applying (1) seat lateral translation, and (2) seat roll rotation around an axis aligned with the head. 
In this work, the dataset of condition (1) is used.
 
\section{Model Posture}

Head posture significantly affects the muscle effort required to maintain an upright head position \cite{caneiro2010, aroeira2017}. 
In individuals without physical impairment, the center of gravity (CG) of the head is typically located approximately $4.3$ to $34.5$ mm anterior to the occipitocervical (OC) spine (e.g., connection of the skull with the upper part of the neck) \cite{Yoganandan2009}. 
While in the axial and coronal plane the head-neck-T1 can be approximated as symmetrical. 
Based on this, the model’s initial pitch posture was defined with the lower neck joint at $0^\circ$ and the upper neck joint at $11.36^\circ$, resulting in a head CG–T1 anterior displacement of $26.5$ mm. 
In the model, the lower neck joint represents the T1–C7 connection, while the upper neck joint corresponds to the atlanto-occipital joint (C0–C1). The selected initial joint angles were not derived from the experimental dataset but were instead chosen to ensure that the head CG–T1 displacement remained within the range reported in the literature for healthy subjects.

\begin{table*}[b]
\centering
\caption{Comparison of MPC cost function weights before and after configuration changes.}
\label{tab:MPC_weights}
\renewcommand{\arraystretch}{1.2}
\begin{tabular}{|l|cccc|}
\toprule
 & $W_{ty1}$ & $W_{ty2}$ & $W_{w_{y1}}$  & $W_{w_{y2}}$  \\
\midrule
 Weights in \cite{Messiou2025}  & 76.96 & 3.37  & 8.26    & 1.62  \\
 New weights  & 78.92 & 15.53 & 15.28 & 41.40 \\
\bottomrule
\end{tabular}
\end{table*}

To identify the model posture after the transient response, the MPC weights for the AP perturbation were re-tuned to reflect the updated MPC configuration. 
Although the AP perturbations optimized the weights for AP head-neck dynamics, the resulting posture influences the lateral dynamic response due to the anterior displacement of the head’s CG, which was not explicitly tuned.
So, AP weights were re-tuned, using upper and lower bounds based on the Pareto-optimal solutions previously identified \cite{Messiou2025}, but under lateral perturbations to account for the posture. 
The re-optimized weights showed a minor adjustment for $W_{ty1}$ while $W_{ty2}$ increased by approximately a factor of $\approx 4.6$, $W_{w_{y1}}$ by a factor of $\approx 1.8$ and $W_{w_{y1}}$ by a factor of $\approx 25.6$ (Table \ref{tab:MPC_weights}).
The high-level optimization method remained unchanged \cite{Messiou2025}.

\begin{figure}[t]
    \centering
    \subfloat[\centering ]{{\includegraphics[width=0.5\linewidth]{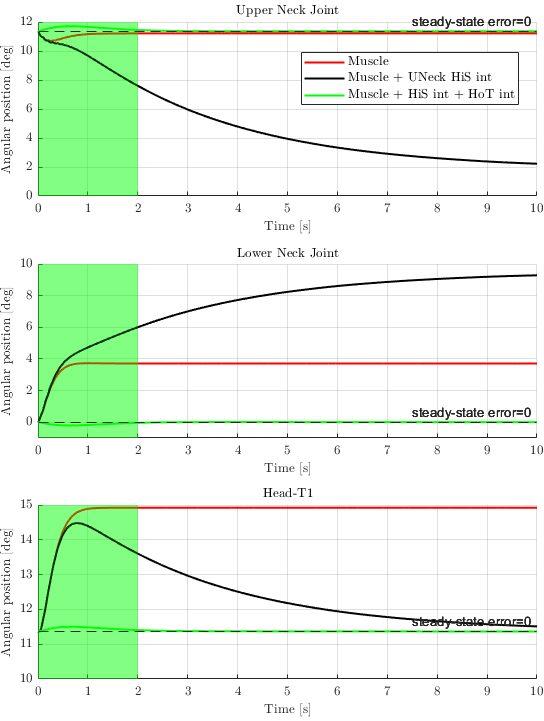} }}%
    \qquad
    \subfloat[\centering ]{{\includegraphics[width=0.42\linewidth]{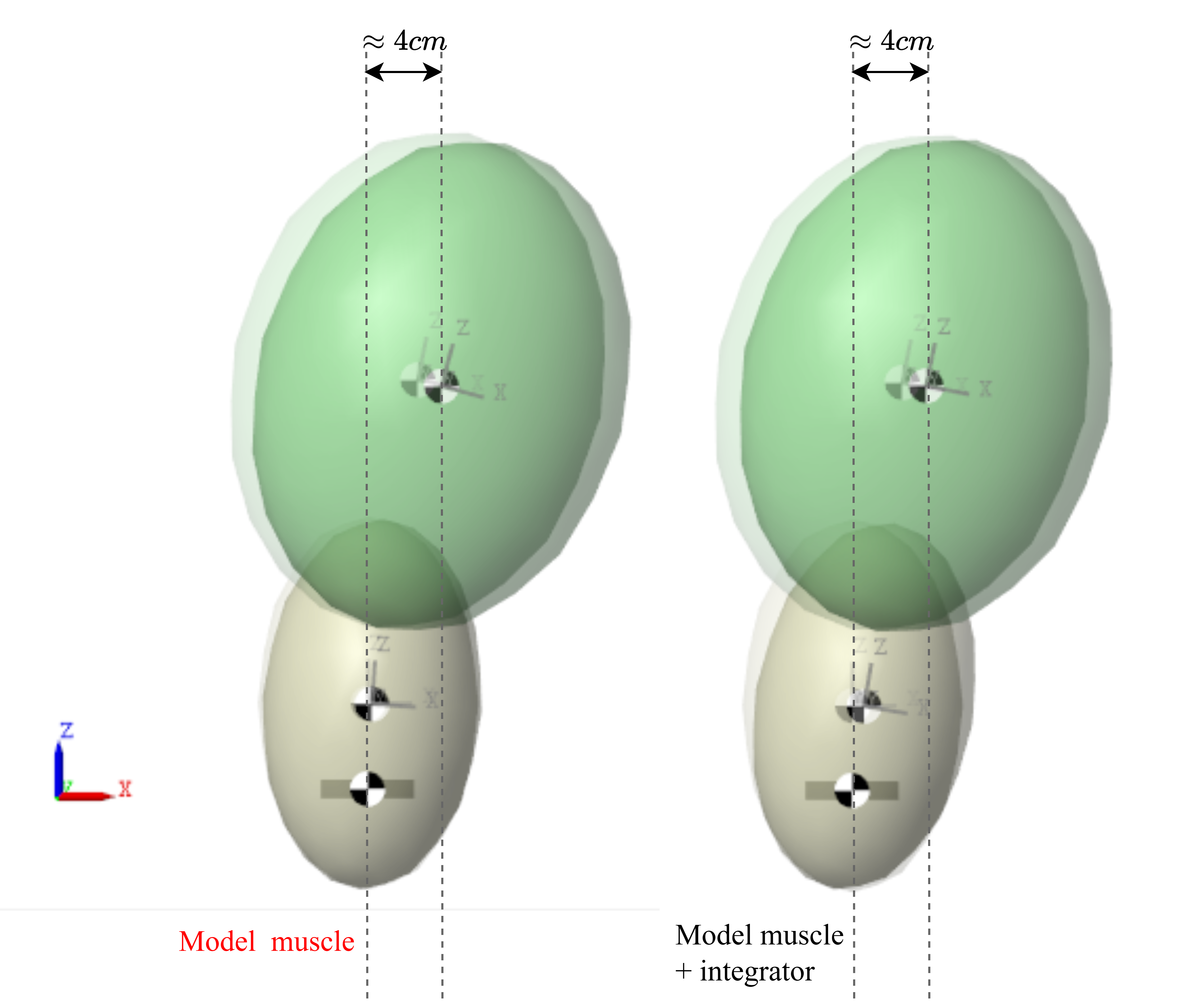} }}%
    \caption{(a) Transient and steady-state responses for three model configurations: (1) HiS and HoT integrators at both joints; (2) muscle effort with partial somatosensory feedback and a HiS integrator at the upper neck joint; (3) muscle effort with partial somatosensory feedback (no integrators). 
    (b) Final postures for configuration (2) and (3).}%
    \label{fig:posture}%
\end{figure}

The model was then initialized without external perturbations to assess the steady-state response and any residual steady-state error. 
The model with muscle effort and partial somatosensory feedback reached steady-state within approximately $0.8$ s (Muscle, Figure \ref{fig:posture}). 
However, the lower neck joint exhibited a steady-state error of approximately $4^\circ$, resulting in an overall head–T1 angle error of about $4.5^\circ$ and a head CG–T1 displacement of $39.86$ mm. 
This displacement slightly exceeded the upper bound reported for healthy subjects ($34.5$ mm).

To prevent drift from the selected starting position, head-in-space (HiS) and head-on-trunk (HoT) integrators were added to both the upper and lower neck joints, following the approach in \cite{Desai2024}. 
The HiS integrator provides additional torque to each joint based on the integral of the global head-in-space angle, while the HoT integrator applies torque based on the integral of the relative joint angle.
The activation of the HiS and HoT integrators effectively eliminated the steady-state error in the model (Figure~\ref{fig:posture}).
However, when evaluating the dynamic response under AP perturbation, this configuration produced unrealistic behavior, significantly deteriorating the accuracy of the model's dynamic response.

By selectively deactivating the HiS and HoT integrators one by one at the upper and lower neck joints and manually adjusting the configuration, a model producing a realistic dynamic response under the AP perturbation was achieved. 
The final configuration combined muscle effort and partial somatosensory feedback with a HiS integrator active only on the upper neck joint (Muscle + UNeck HiS int, Fig.~\ref{fig:posture}).
In this configuration , the head–T1 error and the upper neck joint angle approached zero, while the lower neck joint error increased to approximately $10^\circ$. 
Fig. \ref{fig:posture}(b) compares the model posture after the transient response for both the model Muscle and model Muscle + UNeck int configurations. 
Notably, both configurations resulted in a head CG–T1 displacement of approximately $39.86$ mm.
The performance of these configurations during lateral perturbations will be compared in the Results section.

\section{Results}

This section compares the model responses from the two configurations presented in Fig.~\ref{fig:posture} against experimental human data in both the time and frequency domains.
The focus is on accuracy and real time factor (RTF) performance.

The optimal MPC weight vector obtained from the high-level optimization is:

\begin{equation}
    W = \begin{bmatrix}
    W_{tx1} \\
    W_{ty1} \\
    W_{tx2} \\
    W_{ty2} \\
    W_{tz2} \\
    W_{wx1} \\
    W_{wy1} \\
    W_{wx2} \\
    W_{wy2} \\
    W_{wz2} \\
    \end{bmatrix}
    = 
    \begin{bmatrix}
    17.68 \\
    78.92 \\
    63.77 \\
    15.53 \\
    33.90 \\
    70.84 \\
    15.28 \\
    7.85 \\
    41.40 \\
    5.17 \\
    \end{bmatrix}
\end{equation}

\noindent where \( W \) is the MPC weight vector, consisting of muscle effort costs: 
\( W_{tx1} \) (lower neck joint torque about the x-axis), \( W_{ty1} \) (lower neck joint torque about the y-axis), \( W_{tx2} \) (upper neck joint torque about the x-axis), \( W_{ty2} \) (upper neck joint torque about the y-axis), and \( W_{tz2} \) (upper neck joint torque about the z-axis); and somatosensory conflict costs: \( W_{wx1} \) (lower neck joint rate of orientation about the x-axis), \( W_{wy1} \) (lower neck joint rate of orientation about the y-axis), \( W_{wx2} \) (upper neck joint rate of orientation about the x-axis), \( W_{wy2} \) (upper neck joint rate of orientation about the y-axis), and \( W_{wz2} \) (upper neck joint rate of orientation about the z-axis).

The optimized weights indicate that the largest muscle effort weight is assigned to the lower neck joint torque about the y-axis (pitch), suggesting that the solver prioritized minimizing this torque to reduce muscle effort. 
This is consistent with the steady-state error observed in the model ``muscle" configuration from the posture analysis, where a residual error of approximately $4^\circ$ was present. 
In contrast, the smallest muscle effort weight was assigned to the upper neck joint torque about the x-axis, with a ratio of $W_{ty1}/W{ty2} \approx 5$),  indicating lower sensitivity in compensating forces at the upper neck joint in the x-direction. Additionally, the ratio, $W_{tx1}/W{tx2} \approx 0.5$ shows that $W_{tx1}$, the second smallest muscle effort weight, reflects reduced sensitivity in the lower neck joint torque for roll.
Regarding the minimization of the sensory conflict (the `surprise factor'), the highest weight among the somatosensory feedback components was assigned to \( W_{wx1} \), emphasizing the 
CNS's priority in reducing the lower neck joint roll rate of orientation error.

\begin{table*}[t]
\caption{Feature importance table mapping Fevals with MPC weights.}
\label{tab:feature_importance}
\centering
\begin{tabular}{lcccccccccc}
\hline
Feature & $W_{wx1}$ & $W_{wx2}$ & $W_{tx1}$ & $W_{tx2}$ & $W_{wz2}$ & $W_{tz2}$ & $W_{ty1}$ & $W_{ty2}$ & $W_{wy1}$ & $W_{wy2}$ \\
\hline
roll & \cellcolor[HTML]{C00000}3.7374 & \cellcolor[HTML]{FFC000}2.0898 & \cellcolor[HTML]{C00000}3.8626 & \cellcolor[HTML]{5B9BD5}1.4986 & \cellcolor[HTML]{5B9BD5}1.2363 & \cellcolor[HTML]{5B9BD5}1.2852 & \cellcolor[HTML]{ED7D31}3.4240 & \cellcolor[HTML]{FFC000}1.5936 & \cellcolor[HTML]{C00000}4.5244 & \cellcolor[HTML]{FFC000}1.6516 \\
yaw & \cellcolor[HTML]{FFC000}2.4212 & \cellcolor[HTML]{C00000}3.6855 & \cellcolor[HTML]{C00000}5.4965 & \cellcolor[HTML]{5B9BD5}1.4339 & \cellcolor[HTML]{5B9BD5}1.2091 & \cellcolor[HTML]{5B9BD5}1.3550 & \cellcolor[HTML]{ED7D31}2.9482 & \cellcolor[HTML]{ED7D31}3.1401 & \cellcolor[HTML]{C00000}4.7159 & \cellcolor[HTML]{ED7D31}2.6410 \\
y & \cellcolor[HTML]{5B9BD5}1.4253 & \cellcolor[HTML]{C00000}4.3852 & \cellcolor[HTML]{FFC000}2.0510 & \cellcolor[HTML]{5B9BD5}1.3166 & \cellcolor[HTML]{5B9BD5}1.2004 & \cellcolor[HTML]{5B9BD5}1.0075 & \cellcolor[HTML]{C00000}4.0758 & \cellcolor[HTML]{FFC000}1.7037 & \cellcolor[HTML]{ED7D31}2.5039 & \cellcolor[HTML]{FFC000}2.0768 \\
wroll & \cellcolor[HTML]{C00000}4.6526 & \cellcolor[HTML]{C00000}6.1938 & \cellcolor[HTML]{FFC000}2.0109 & \cellcolor[HTML]{FFC000}2.5443 & \cellcolor[HTML]{FFC000}2.0170 & \cellcolor[HTML]{5B9BD5}1.2523 & \cellcolor[HTML]{ED7D31}2.7351 & \cellcolor[HTML]{FFC000}2.0365 & \cellcolor[HTML]{FFC000}2.3173 & \cellcolor[HTML]{FFC000}1.7733 \\
wyaw & \cellcolor[HTML]{FFC000}1.9621 & \cellcolor[HTML]{FFC000}2.1620 & \cellcolor[HTML]{FFC000}2.4303 & \cellcolor[HTML]{5B9BD5}1.2559 & \cellcolor[HTML]{FFC000}2.2882 & \cellcolor[HTML]{5B9BD5}1.3042 & \cellcolor[HTML]{FFC000}2.4438 & \cellcolor[HTML]{FFC000}1.9926 & \cellcolor[HTML]{ED7D31}3.3265 & \cellcolor[HTML]{FFC000}1.6553 \\
vy & \cellcolor[HTML]{ED7D31}3.2837 & \cellcolor[HTML]{FFC000}2.2705 & \cellcolor[HTML]{ED7D31}3.1693 & \cellcolor[HTML]{ED7D31}2.7131 & \cellcolor[HTML]{FFC000}1.8131 & \cellcolor[HTML]{5B9BD5}1.4763 & \cellcolor[HTML]{ED7D31}2.6310 & \cellcolor[HTML]{FFC000}1.7409 & \cellcolor[HTML]{FFC000}2.1529 & \cellcolor[HTML]{FFC000}1.6098 \\
$\|f_{vy}\|$ & \cellcolor[HTML]{C00000}4.3099 & \cellcolor[HTML]{ED7D31}3.2040 & \cellcolor[HTML]{FFC000}1.8622 & \cellcolor[HTML]{5B9BD5}1.3760 & \cellcolor[HTML]{5B9BD5}1.2164 & \cellcolor[HTML]{5B9BD5}1.1012 & \cellcolor[HTML]{C00000}3.9552 & \cellcolor[HTML]{FFC000}1.7236 & \cellcolor[HTML]{FFC000}2.1230 & \cellcolor[HTML]{ED7D31}3.1200 \\
$\|f_{wroll}\|$ & \cellcolor[HTML]{ED7D31}3.1472 & \cellcolor[HTML]{C00000}3.8092 & \cellcolor[HTML]{ED7D31}3.4800 & \cellcolor[HTML]{FFC000}1.9024 & \cellcolor[HTML]{FFC000}1.5639 & \cellcolor[HTML]{5B9BD5}1.1975 & \cellcolor[HTML]{ED7D31}2.7295 & \cellcolor[HTML]{5B9BD5}1.4829 & \cellcolor[HTML]{FFC000}1.9584 & \cellcolor[HTML]{FFC000}1.9172 \\
$\|f_{wyaw}\|$ & \cellcolor[HTML]{5B9BD5}1.3595 & \cellcolor[HTML]{C00000}4.0266 & \cellcolor[HTML]{ED7D31}2.7426 & \cellcolor[HTML]{5B9BD5}1.0785 & \cellcolor[HTML]{5B9BD5}1.3150 & \cellcolor[HTML]{5B9BD5}1.1453 & \cellcolor[HTML]{FFC000}2.3658 & \cellcolor[HTML]{FFC000}2.2939 & \cellcolor[HTML]{ED7D31}2.8097 & \cellcolor[HTML]{FFC000}1.6024 \\
$\phase f_{vy}$ & \cellcolor[HTML]{C00000}5.9543 & \cellcolor[HTML]{C00000}4.2117 & \cellcolor[HTML]{ED7D31}3.0237 & \cellcolor[HTML]{FFC000}1.5447 & \cellcolor[HTML]{5B9BD5}1.2832 & \cellcolor[HTML]{5B9BD5}1.1801 & \cellcolor[HTML]{ED7D31}2.5618 & \cellcolor[HTML]{FFC000}1.8118 & \cellcolor[HTML]{FFC000}2.1057 & \cellcolor[HTML]{FFC000}2.3863 \\
$\phase f_{wroll}$ & \cellcolor[HTML]{5B9BD5}0.5784 & \cellcolor[HTML]{5B9BD5}1.0841 & \cellcolor[HTML]{5B9BD5}0.6865 & \cellcolor[HTML]{5B9BD5}0.4038 & \cellcolor[HTML]{5B9BD5}0.4781 & \cellcolor[HTML]{5B9BD5}0.3637 & \cellcolor[HTML]{5B9BD5}0.8058 & \cellcolor[HTML]{5B9BD5}0.5233 & \cellcolor[HTML]{5B9BD5}0.5550 & \cellcolor[HTML]{5B9BD5}0.8403 \\
$\phase f_{wyaw}$ & \cellcolor[HTML]{5B9BD5}1.4946 & \cellcolor[HTML]{FFC000}2.0418 & \cellcolor[HTML]{ED7D31}2.6473 & \cellcolor[HTML]{5B9BD5}1.3906 & \cellcolor[HTML]{5B9BD5}1.1615 & \cellcolor[HTML]{5B9BD5}0.9959 & \cellcolor[HTML]{FFC000}1.7390 & \cellcolor[HTML]{FFC000}1.7259 & \cellcolor[HTML]{FFC000}1.9010 & \cellcolor[HTML]{5B9BD5}1.2808 \\
\hline
\end{tabular}
\end{table*}

To further assess the relative impact of these weights, a feature importance analysis was performed using the random forest algorithm. 
By collecting MPC weight sets and corresponding objective function evaluations (Fevals) during the high-level optimization process, a random forest model was trained to quantify the importance of each weight across all evaluation functions (Table~\ref{tab:feature_importance}).
The analysis revealed that among the muscle effort terms, \( W_{tx1} \) (lower neck torque about x) and \( W_{ty1} \) (lower neck torque about y) had the greatest influence on model performance, underscoring the critical role of lower neck joint control in achieving accurate dynamic responses. 
For the somatosensory error components, both \( W_{wx1} \) and \( W_{wx2} \) (lower and upper neck roll rates of orientation, respectively) were highly influential, particularly for the y-axis translational and rotational metrics, which is consistent with the direction of the lateral perturbation.
Additionally, the somatosensory feedback from the lower neck joint \( W_{wy1} \) (lower neck pitch rate of orientation) was found to be especially important in roll and yaw dynamics.
This further supports the significance of lower joint control and suggests that the anterior positioning of the head's CG influences the coupled roll and yaw responses.

 \begin{figure}[t]
  \centering 
  \includegraphics[width=1\linewidth] {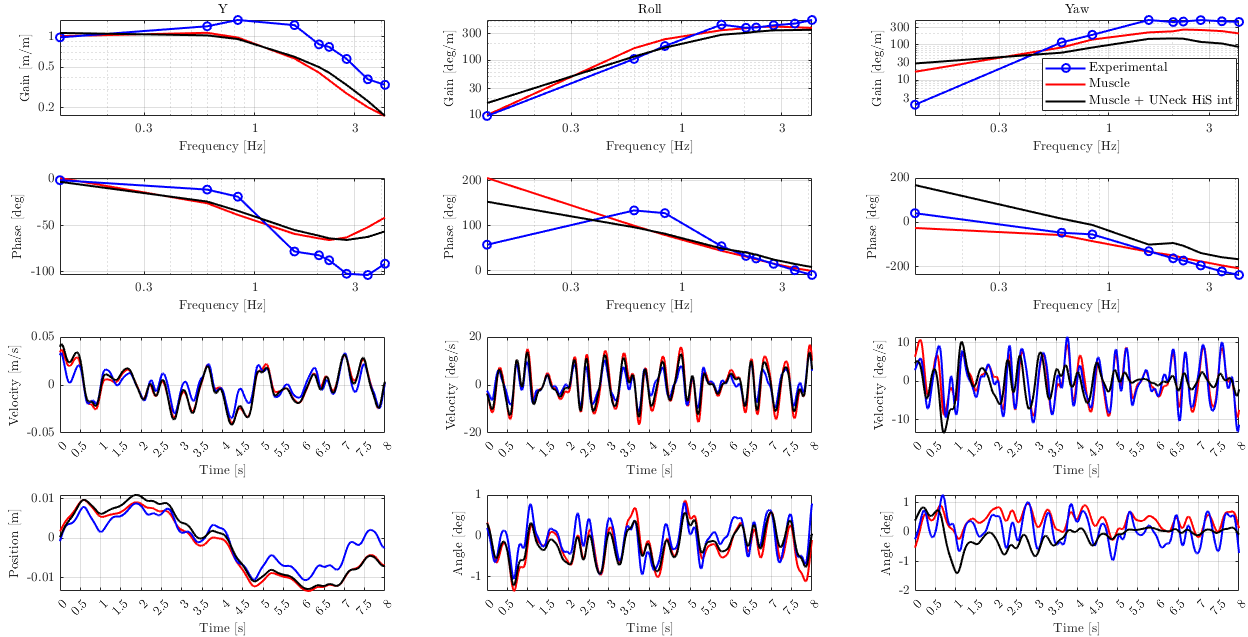}
  \caption{Dynamic response of the head-neck model under lateral multi-sine perturbations. Comparison includes: (1) muscle effort with partial somatosensory feedback; (2) same configuration with a HiS integrator at the upper neck joint; (3) experimental human data. Results are shown in both time and frequency domains.}
  \label{fig:HeadNeckModel}
\end{figure}

Figure~\ref{fig:HeadNeckModel} presents the dynamic response of the head-neck model under lateral multi-sine perturbations. The comparison includes both time and frequency domain analyses for the following configurations:
(1) Muscle: muscle effort combined with partial somatosensory feedback;
(2) Muscle + UNeck HiS int: the same configuration as (1) with the addition of a head-in-space (HiS) integrator at the upper neck joint, and
(3) Experimental human data.
The results indicate that the muscle configuration generally provides a closer match to the experimental responses in both the time and frequency domains. 
The only exception is observed at lower frequencies in the yaw response, where the muscle + UNeck HiS int configuration shows a slightly improved gain alignment with the experimental data.
Furthermore, the RTF during simulations was approximately 8-11, measured on a Windows 10 desktop equipped with an Intel\textregistered{} Xeon\textregistered{} W-2133 CPU (3.60~GHz base clock speed, 6 physical cores, 12 logical processors) and 64.0~GB of RAM.

\section{Conclusion}

The results indicate that the configuration using muscle effort combined with partial somatosensory feedback provides the best overall dynamic response during lateral perturbations. 
Although this configuration exhibits the highest steady-state error, this can be attributed to the selected initial posture, which was not derived from experimental data but was chosen to remain within the head CG–T1 anterior displacement range reported in the literature. 
The final displacement slightly exceeded this range, likely due to the inherent simplifications of the Simscape biomechanical model. 
Nevertheless, the model effectively reproduces postural stabilization without the need for integrators.
This demonstrates that muscle effort and partial somatosensory feedback are sufficient to capture the CNS decision making during laterally perturbed head-neck dynamics responses, as proven already for anterior-posterior perturbations \cite{Messiou2025}.

\bibliographystyle{spbasic.bst}
\bibliography{00_Main}

\end{document}